\documentclass{elsart3}

\usepackage{natbib}

 \usepackage{graphics}

\usepackage{amssymb}

\begin{document}

\begin{frontmatter}



\title{                                                                        
A Detection of the Integrated Sachs-Wolfe Effect}


\author[HavPrin]{S.P. Boughn} and 
\author[Port]{R.G. Crittenden}

\address[HavPrin]{
Department of Astronomy, Haverford College,
Haverford, PA 19041 \\ Department of Physics, Princeton
University, Princeton, NJ 08544}
\address[Port]{
Institute of Cosmology and Gravitation, University of Portsmouth, Portsmouth
PO1 2EG  UK}
\date{\today}

\begin{abstract}
We have detected statistically significant correlations between the
cosmic microwave background and two tracers of large-scale structure,
the HEAO1 A2 full sky hard X-ray map \citep{Boldt} and the NVSS 1.4 GHz,
nearly full sky radio galaxy survey \citep{Condon}.  The level
of correlations in these maps is consistent with that predicted for the
integrated Sachs-Wolfe (ISW) effect in the context of a $\Lambda CDM$
cosmological model and, therefore, provides independent evidence for a
cosmological constant.  A maximum likelihood fit to the amplitude of the
ISW effect relative to the predicted value is $1.13 \pm 0.35$
(statistical error only).
\end{abstract}


\end{frontmatter}


In the currently favored $\Lambda$-cold dark matter ($\Lambda CDM$) cosmological 
model, the dominant energy content of the universe 
is due to a cosmological constant, $\Lambda$, or some other
form of ``dark energy''.  The primary evidence for this model comes from supernovae 
redshift/magnitude observations \citep[e.g.][]{Barris} that imply the expansion of 
the universe is accelerating and from the spatial power spectrum of the fluctuations in 
the cosmic microwave background \citep{WMAP}.  One of the consequences of this
model is that a significant portion of the anisotropy of the $CMB$ is produced 
recently via a mechanism known as the integrated Sachs-Wolfe (ISW) effect 
\citep{SW}. This anisotropy is created as CMB photons traverse the
evolving gravitational potentials of linear density perturbations (i.e., 
$\delta \rho / \rho \ll 1$) at relatively low redshifts ($z \lesssim 1$).  
Crittenden and Turok (1996) 
suggested that the ISW effect could be detected by 
correlating the CMB with some nearby ($z \lesssim 1$) tracer of matter, e.g., galaxies 
or AGN. 

A significant ISW effect arises recently in a curvature dominated universe or in 
a flat universe if it is $\Lambda$ dominated. 
On the other hand, in a flat, matter dominated universe the gravitational
potentials of linearly collapsing structures are constant and there is no ISW effect.
It is in this sense that the ISW effect has a unique sensitivity to a cosmological 
constant in a flat universe.  If there is a significant $\Lambda$, then the CMB will
be correlated with tracers of mass at low redshift;  if not, there is no such 
correlation.  To be sure, such correlations could be generated at roughly the same level
if the universe is significantly open with $\Omega \lesssim 0.5$; however, the position of 
the Doppler peaks in the CMB anisotropy spectrum measured by WMAP and other experiments 
indicate that this is not the case \citep{WMAP}.

We have cross-correlated two mass tracers with the WMAP 
'internal linear combination' (ILC) CMB map and found
significant correlations in both cases \citep{Nature}.  The
$HEAO1~A2$ 2-10 $keV$ all sky X-ray map provides a good tracer of mass with a 
median redshift of about $z \sim 0.9$ \citep{xray_bias}.  Likewise, 
the NVSS $1.4~GHz$ nearly full sky ($82\%$) radio catalog also provides a good
tracer with the same median redshift \citep{Condon}.  Both of these maps 
have significant systematics which we have tried to identify and correct for 
\citep{BC_PRL,BCK}.  However, even if there are still residual contaminants, 
it is unlikely that the systematics in any of the three disparate data sets are 
correlated.  In fact, the resulting cross-correlation functions are largely
independent of the corrections.

\begin{figure}
\centerline{
\resizebox{3in}{!}{\rotatebox{270}{\includegraphics{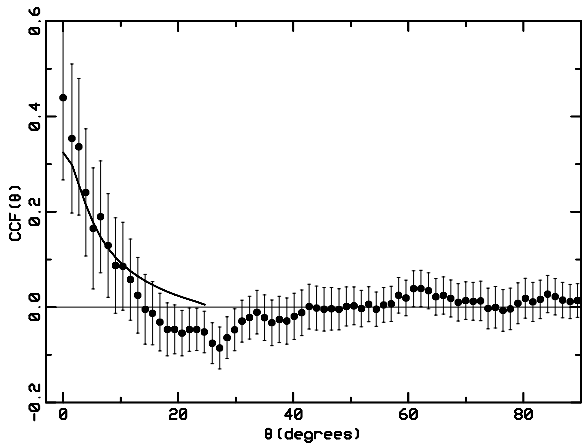}}}}
\caption{
The cross-correlation function of the WMAP ILC CMB maps with the
$HEAO1~A2$ 2-10 $keV$, hard X-ray map.  The error bars were determined from
Monte Carlo calculations and are highly correlated.  The solid curve is the
predicted ISW effect from the standard $\Lambda CDM$ cosmological model and
is not a fit to the data. The units are $\mu K~TOT~counts~s^{-1}$ where 
$1~TOT~count~s^{-1} \sim 1 \times 10^{-5}erg~s^{-1}cm^{-2}sr^{-1}$.
}
\label{fig:heao}
\end{figure}
 
\begin{figure}
\centerline{
\resizebox{3in}{!}{\rotatebox{270}{\includegraphics{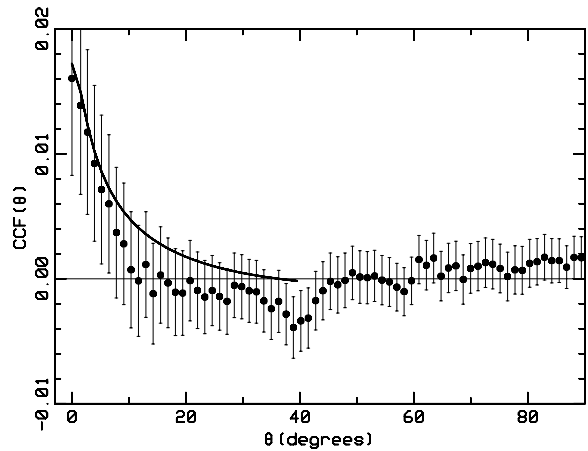}}}}
\caption{
The cross-correlation function of the WMAP ILC CMB maps with the
NVSS 1.4 $GHz$ radio survey.  The error bars were determined from
Monte Carlo calculations and are highly correlated.  The solid curve is the
predicted ISW effect from the standard $\Lambda CDM$ cosmological model and
is not a fit to the data.  The units are $mK~counts$ where the counts
are number of radio sources per $1.3 \times 1.3$ degree pixel.
}
\label{fig:nvss}
\end{figure}
 
A standard measure of the correlation of two data sets is the cross-correlation
function ($CCF$), which in this case is defined by
\begin{equation}
CCF(\theta) = {1 \over N_{\theta}} \sum_{i,j} (S_i -\bar{S})(T_j -\bar{T})
\end{equation}
where the sum is over all pairs of pixels separated by an angle
$\theta$, $S_i$ is the signal strength of the $i^{th}$ pixel of the tracer map,
$\bar{S}$ is the mean signal, $T_i$ is the CMB temperature of the $i^{th}$ pixel,
$\bar{T}$ is the mean CMB temperature, and $N_{\theta}$ is the number of pairs 
of pixels separated by $\theta$.  In the case of the $HEAO$ X-ray map, $S$ is the
2-10 $keV$ X-ray intensity and for the $NVSS$ radio catalog, $S$ is the surface
density of radio sources.  Figure 1 is the X-ray $CCF$ and Figure 2 is the $NVSS$
$CCF$.  In both cases, the error bars were computed from Monte Carlo simulations
and are highly correlated.  These errors are primarily due to the fluctuations 
inherent in the X-ray, radio, and CMB backgrounds with only minor contributions due
instrument noise and this is why they are so highly correlated.  Photon shot noise in
the X-ray map and Poisson noise due to finite source counts are important and 
account for roughly a half of the error.

The solid curves in the two figures are predictions of the currently favored 
$\Lambda CDM$ cosmological model and are not fits to the data.  The statistical 
significance of the detection of the correlation is roughly $2.5~\sigma$ in both
cases and it is clear that the correlation is consistent with the predicted ISW
effect.  Recall that in a flat, matter dominated universe, there is no expected 
correlation.  A maximum likelihood fit to both data sets, taking into account that
the two tracer maps are themselves correlated with each other, yields an ISW 
amplitude of $1.13 \pm 0.35$ relative to the predicted amplitude for the WMAP 
best fit cosmological model.  This implies
a $3.2~\sigma$ detection of the ISW effect with an amplitude 
consistent with that predicted by the standard $\Lambda CDM$ cosmological model.

The predicted ISW effect is sensitive to the biases of the two tracer maps, 
the redshift distribution of the sources, and to the the cosmological model.
The redshift distributions were taken from model luminosity functions  
\citep{DP,Cowie} and the biases were determined from the 
the clustering of the sources in the two tracer maps \citep{BC_PRL,xray_bias},
the source redshift distributions, and the standard
cosmological model.  Even though there is considerable uncertainty in the luminosity
functions and concomitant source redshift distributions, the predicted ISW effect is
relatively insensitive to these uncertainties.  For example, if the redshift 
distribution is artificially low, so will be the implied bias.  However, in a 
$\Lambda CDM$ universe at these redshifts, the intrinsic ISW effect is somewhat 
larger at low redshifts and these two errors tend to cancel each other.  For this 
reason, we conclude that this type of systematic error will be considerably smaller 
than the statistical errors indicated in Figures 1 and 2.  

Another possible systematic error in the case of the X-ray/CMB analysis is nearby,
unresolved source contamination if such sources emit significantly in both X-rays
and microwaves.  We masked the CMB map with the most aggressive WMAP mask
($k0$) that includes both the plane of the Galaxy and high latitude sources.  The
Galactic plane was also removed from the X-ray and, in addition, we masked high 
Galactic latitude X-ray sources in a variety of ways.   However, the X-ray CCF is
largely independent of how these sources are masked.  

The NVSS/CMB CCF is less 
sensitive to high Galactic latitude sources since it is based on radio number 
counts rather than source flux.  More insidious is microwave emission from distant 
X-ray and radio sources since it is these sources that are purported to map the 
mass distribution that generates the ISW signal.  Models of AGN and radio sources 
indicate that they are not a source of contamination but an even stronger 
constraint can be obtained from the different WMAP CMB frequency bands 
(41, 61, and 94 $GHz$) maps.  Most radio/microwave sources have spectral indices, 
$\alpha$, ranging from $\alpha \simeq -0.7$ for a synchrotron spectrum to $\alpha \simeq -0.1$ 
for a Bremsstrahlung spectrum.  On the other hand, the blackbody spectral 
index of the CMB is $\alpha \simeq +2.0$ in the Rayleigh-Jeans part of the spectrum.  If 
the CCF we observe were due to radio source contamination then one would expect the 
CCF with the $94~GHz$ WMAP map to be from 5.7 to 9.4 times larger than the CCF with 
the $41~GHz$ map.  Even inverted spectrum sources with spectral indices as large as 
$\alpha \simeq +1.2$ would imply a factor of two difference between these two CCFs. 

\begin{figure}
\centerline{
\resizebox{3in}{!}{\rotatebox{270}{\includegraphics{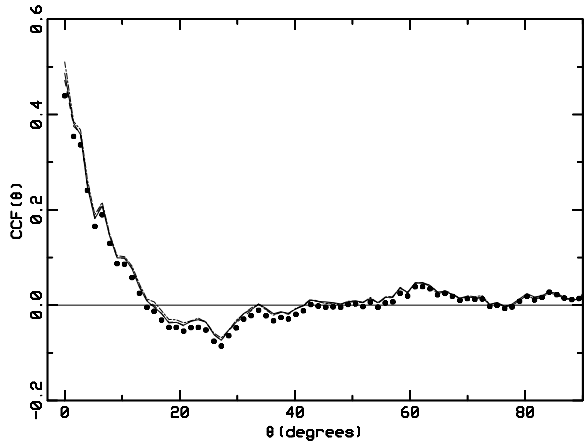}}}}
\caption{
The cross-correlation function of the $HEAO$ X-ray map
with the 41 (solid curve), 61 (dashed curve) and 94 $GHz$ (dot-dashed
curve) WMAP CMB maps.  The points are the cross-correlation with 
the internal linear combination map plotted in Figure 1.
}
\label{fig:heao2}
\end{figure}
 
\begin{figure}
\centerline{
\resizebox{3in}{!}{\rotatebox{270}{\includegraphics{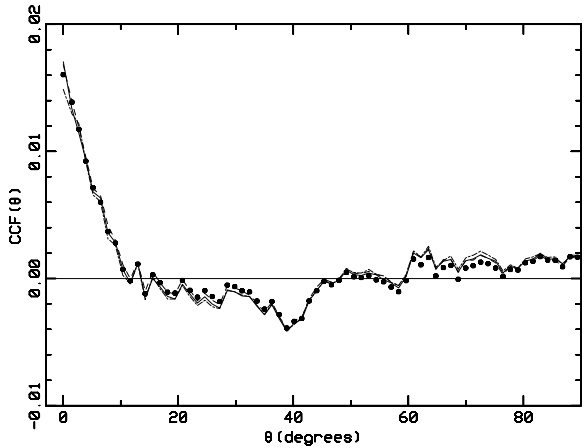}}}}
\caption{
The cross-correlation function of the NVSS radio map
with the 41 (solid curve), 61 (dashed curve) and 94 $GHz$ (dot-dashed
curve) WMAP CMB maps.  The points are the same as those plotted in Figure 2.
}
\label{fig:nvss2}
\end{figure}

Figures 3 and 4 are CCFs of the X-ray and NVSS maps with the 41, 61, and 94 $GHz$ 
CMB maps.  (These three maps were corrected for Galactic emission using the
synchrotron, free-free, and dust maps from the WMAP data public data set.)
The solid and two dashed lines in the figures indicate the three CCFs while the 
points are the CCFs from Figures 1 and 2.  It is clear that the difference 
between them is a few percent at most and we conclude that it is extremely unlikely 
that the observed correlation is due to radio source contamination.

The correlation functions of Figures 1 and 2 are consistent with the ISW effect predicted
in a universe dominated by a cosmological constant, $\Lambda \sim 0.7$ and, in this
sense, provide important 
independent confirmation of the currently favored $\Lambda CDM$ cosmological 
model.

We are grateful to M. Nolta, L. Page and the rest of the WMAP team as well as N. Turok.
We acknowledge the use of the Legacy Archive for Microwave Background Data Analysis 
(LAMBDA). Support for LAMBDA is provided by the NASA Office of Space Science



\end{document}